\documentclass[aps,reprint,prc,superscriptaddress]{revtex4-1}

\usepackage{CJK}

\usepackage{graphicx}
\usepackage{dcolumn}
\usepackage{bm}

\usepackage[utf8]{inputenc}
\usepackage{amsmath}
\usepackage{amsfonts}
\usepackage{mathrsfs}
\usepackage{amssymb}
\usepackage{xfrac}
\usepackage{graphicx}
\usepackage{enumitem}
\usepackage{booktabs}
\usepackage{hyperref}
\usepackage{setspace}
\usepackage{url}

\usepackage{xcolor}
\usepackage{etoolbox}
\apptocmd{\sloppy}{\hbadness 10000\relax}{}{}
\hypersetup{
	colorlinks,
	linkcolor={black},
	citecolor={blue!50!black},
	urlcolor={blue!80!black}
}

\AtBeginDocument{
	\heavyrulewidth=.08em
	\lightrulewidth=.05em
	\cmidrulewidth=.03em
	\belowrulesep=.65ex
	\belowbottomsep=0pt
	\aboverulesep=.4ex
	\abovetopsep=0pt
	\cmidrulesep=\doublerulesep
	\cmidrulekern=.5em
	\defaultaddspace=.5em
}

\makeatletter
\def\@author#1{\g@addto@macro\elsauthors{\normalsize%
		\def\baselinestretch{1}%
		\upshape\authorsep#1\unskip\textsuperscript{%
			\ifx\@fnmark\@empty\else\unskip\sep\@fnmark\let\sep=,\fi
			\ifx\@corref\@empty\else\unskip\sep\@corref\let\sep=,\fi
		}%
		\def\authorsep{\unskip,\space}%
		\global\let\@fnmark\@empty
		\global\let\@corref\@empty  
		\global\let\sep\@empty}%
	\@eadauthor={#1}
}
\makeatother

\begin{document}

\begin{CJK*}{GB}{} 

\title{Comparison of the Structure Function $F_2$ as Measured by Charged Lepton and Neutrino Scattering from Iron Targets} 





\author{N.~Kalantarians}
     \email{narbe@jlab.org}
\affiliation{Virginia Union University, Richmond, VA 23220 }

\author{C.~Keppel}
\affiliation{Thomas Jefferson National Accelerator Facility, Newport News, Virginia 23606}

\author{M.E.~Christy}
\affiliation{Hampton University, Hampton, Virginia 23668}
\affiliation{Thomas Jefferson National Accelerator Facility, Newport News, Virginia 23606}


\date{\today}


\begin{abstract}
A comparison study of world data for the structure function $F_2$ for Iron, as measured by both charged lepton and neutrino scattering experiments, is presented. Consistency of results for both charged lepton and neutrino scattering is observed for the full global data set in the valence regime. Consistency is also observed at low $x$ for the various neutrino data sets, as well as for the charged lepton data sets, independently. However, data from the two probes exhibit differences on the order of 15\% in the shadowing/anti-shadowing transition region where the Bjorken scaling variable $x$ is $<$ 0.15. This observation is indicative that neutrino probes of nucleon structure might be sensitive to different nuclear effects than charged lepton probes. Details and results of the data comparison are here presented.  
\end{abstract}


\maketitle

\end{CJK*}


\section{Introduction}

A complete and fundamental understanding of nucleon and nuclear structure in terms of the underlying partonic constituents is one of the outstanding challenges in hadron physics today. High energy lepton scattering provides one of the most powerful tools to investigate this structure. In this process, contributions to the measured nucleon structure function $F_2$ can be expressed in terms of the parton distribution functions (PDFs) of the nucleon. Interestingly, comparisons of lepton scattering from various nuclear target $F_2$ data display nuclear medium modifications as demonstrated by the measured structure function ratios of heavy nuclei to the deuteron, $F_2^A/F_2^D$, first noted famously by the European Muon Collaboration (EMC). The behavior of this ratio has since been broadly divided into four regions: $x \lesssim 0.1$ the shadowing region; $0.1 \lesssim x \lesssim 0.3$ the anti-shadowing region; $0.3 \lesssim x \lesssim 0.8$, the EMC effect region; and greater than $x \approx 0.8$ the Fermi motion region. Many analyses have been performed to study this complex behavior, and several global phenomenological parameterizations for nuclear parton distribution functions (NPDFs) have been developed which successfully reproduce the nuclear modifications to lepton-nucleon scattering ~\cite{Hirai, EPS, SK}.

It has been observed through such global NPDF fitting efforts ~\cite{SK1, SK2, SK3} that the $F_2^A/F_2^D$ ratio may be different between charged lepton and neutrino scattering data. Neutrino scattering data have long been predicted to display for instance more shadowing ~\cite{QPred}, with explanations spanning off-shell effects ~\cite{KP}, charge symmetry effects ~\cite{BLT}, meson cloud contributions 
~\cite{NME}, interference amplitudes from multiple scattering of quarks ~\cite{Stan}, and beyond ~\cite{BK,KP2,KP3}. It has also been suggested that there is not yet any verified difference, but an observation derived rather from the use of a particular NPDF fitting approach ~\cite{HPCS}. The question as to whether there is some probe dependence to the observed structure function has, therefore, remained something of a puzzle. Furthermore, relative to charged lepton scattering, the experimental evidence for shadowing in neutrino scattering is scant and comparatively new ~\cite{KMS}.  

In all, it is important to note that neutrino $F_2^D$ have been constructed from PDF's, which are parameterized from charged lepton data, due to the paucity of available neutrino-deuteron scattering data. In this paper we provide a deuteron model-independent comparison of world data for the structure function as measured from Iron targets only, using both charged lepton and neutrino scattering probes, for the purpose of testing how large a role the constructed neutrino $F_2^D$ has actually has played in comparisons between charged lepton and neutrino data sets. All data employed in this study are in the deep inelastic scattering (DIS) region of four-momentum transfer $Q^2>2$ GeV$^2$ and final state invariant mass 
$W^2>4$ GeV$^2$, and cover a Bjorken scaling variable $x$ range where the EMC effect, shadowing, and anti-shadowing regimes reside. We have chosen Iron as it is the only nucleus for which the latter broad range of kinematic coverage is available from both neutrino and charged lepton scattering experiments. 

We stress that this is purely a comparative analysis of existing charged lepton and neutrino scattering data. The data have had few and small, if any, corrections applied beyond what was published originally by each respective collaboration. One observation of this combined analysis is the consistency of the global data set for the charged lepton and neutrino results at larger $x$. Consistency is also observed at low $x$ for the various neutrino data sets, as well as for the charged lepton data sets, independently.


\section{The data sets}

\begin{figure}{}
\includegraphics[width=8cm,height=7cm]{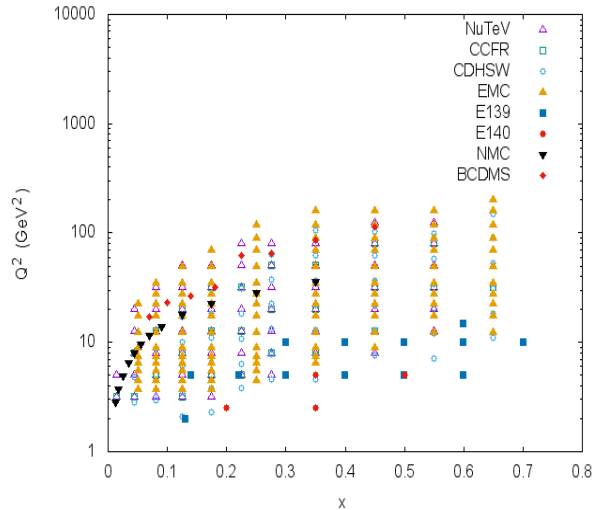}
\caption{\label{F2data_Q2vsx} (Color online) Scatter plot of available $F_2^{Fe}$ data in $x$, $Q^2$ kinematics with the conventional deep inelastic scattering cuts applied. Charged lepton data are denoted by solid symbols, while neutrino data are denoted with open symbols.}
\end{figure}

The phase space plot in Figure ~\ref{F2data_Q2vsx} illustrates the data in $Q^2$ and $x$ used in this analysis. The majority of the data are available from the online Durham HepData Project Database ~\cite{Durham1,Durham2}. The neutrino (and anti-neutrino) structure function data sets employed in this study are NuTeV ~\cite{NUTEV}, CDHSW ~\cite{CDHSW}, and CCFR ~\cite{CCFR}, as provided at the database. The charged lepton BCDMS ~\cite{BCDMS} and NMC ~\cite{NMC1,NMC2} experiment data sets are available at the database in structure function ratios of Iron to Deuteron. For these two, a reliable parameterization from the NMC collaboration ~\cite{F2ALLM} of the deuteron structure function $F_2^D$, fit over a broad kinematic DIS region, was used to extract $F_2^{Fe}$ multiplicatively. The use of this parameterization could induce an additional uncertainty to the data of $\approx2\%$. Beyond the database, SLAC experiment E139 electron data ~\cite{E139} were obtained from the E139 web-site ~\cite{E139web}, in the form of inclusive cross-sections, and converted to $F_2$ using a parameterization ~\cite{Christy} for the longitudinal cross section ratio $R = \sigma_{L}/\sigma_{T}$. 




\noindent This extraction of $F_2$ was not assumed to introduce any additional uncertainty to the data, as $R$ is not typically large and moreover the parameterization is well constrained in this region. To study only data in or near the conventional deep inelastic scattering region, kinematic cuts of $Q^2>2$ GeV$^2$ and final state invariant mass $W^2>4$ GeV$^2$ were applied to all of the data sets.

The $F_2^{Fe}$ data were subsequently brought to a common $Q^2$ via $F_2^{allm}(x,Q^2_{common})/F_2^{allm}(x,Q^2_{data}) \times F_2^{Fe}$, where $F_2^{allm}$ is the aforementioned NMC parameterization of $F_2^D$. This parameterization provides an option to utilize the neutron to proton ratio $F_2^n/F_2^p$ with or without $Q^2$ dependence. Both cases were investigated, with negligible difference and we here employ the $Q^2$ dependent version. In order to study uncertainty from the choice of parameterization used in this process, we constructed EMC-effect type ratios, $F_2^{Fe}/F_2^D$, for varying ranges of $Q^2$ values being brought to a common, central $Q^2$, and verified that a consistent EMC-like ratio held, with the $Q^2$ centering dependence less than $2\%$. 

All of the data were isoscalar corrected when published, and no change was made in this work to the published corrections. These corrections were on the order of a few \% for all experiments, as Iron-56 provides a near isoscalar heavy nuclear target. 

In this paper, the errors shown on the data are statistical only. The systematic errors can be found in the individual data publications, and are typically less than 10\% for the neutrino, and less than 5\% for the charged leption, data. The EMC data have a normalization correction of 7\% applied to them, following the global analysis work of Whitlow ~\cite{Whitlow}.  


\section{Results and Discussion}

To compare the charged lepton and neutrino data, the latter were scaled by a factor of 18/5, derived from current algebra to account for the quark charge. At leading order and assuming isospin invariance with no charge symmetry violation in the nucleon, the $F_2$ nucleon structure function probed via charged lepton scattering can be written naively in terms of the u and d (anti)quark distribution functions as

\begin{equation}
\label{eq:F2en}
F^{N}_2(x) = x\frac{5}{18}\left[u(x)+\overline{u}(x)+d(x)+\overline{d}(x)\right].
\end{equation}
\noindent Here, the variables with the bar over them represent the corresponding anti-quarks.
Since neutrinos do not couple to quark charges, the corresponding equation for the $F_2$ nucleon structure function probed via neutrino scattering can be similarly naively written as

\begin{equation}
\label{eq:F2nun}
F^{\nu N}_2(x) = x\left[u(x)+\overline{u}(x)+d(x)+\overline{d}(x)\right].
\end{equation}
\noindent Early data comparing charged lepton and neutrino scattering via these equations was used originally to confirm the fractional charge assignments for the quarks. 

\begin{figure}
\includegraphics[width=8cm,height=7cm]{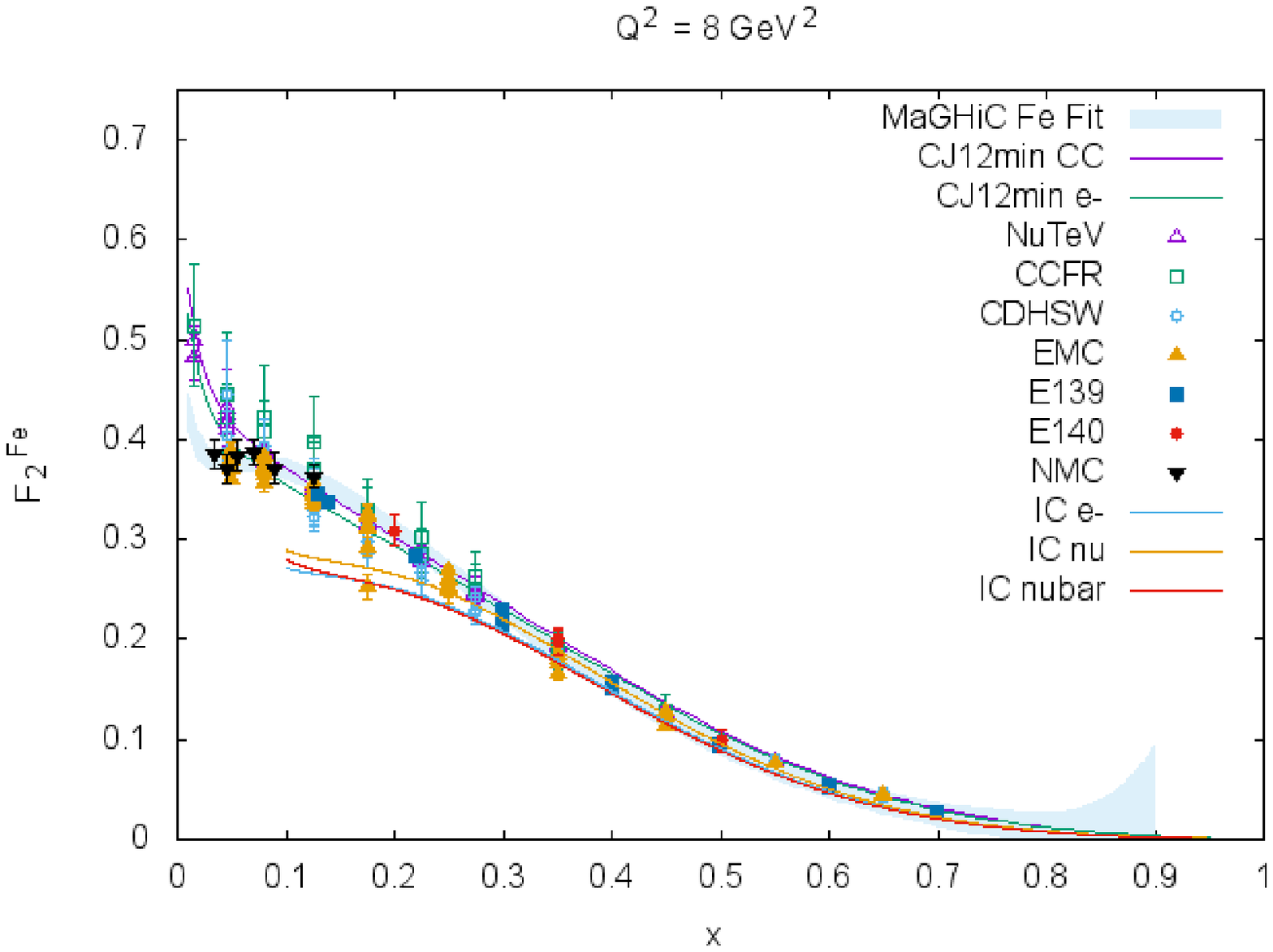}
\includegraphics[width=8cm,height=7cm]{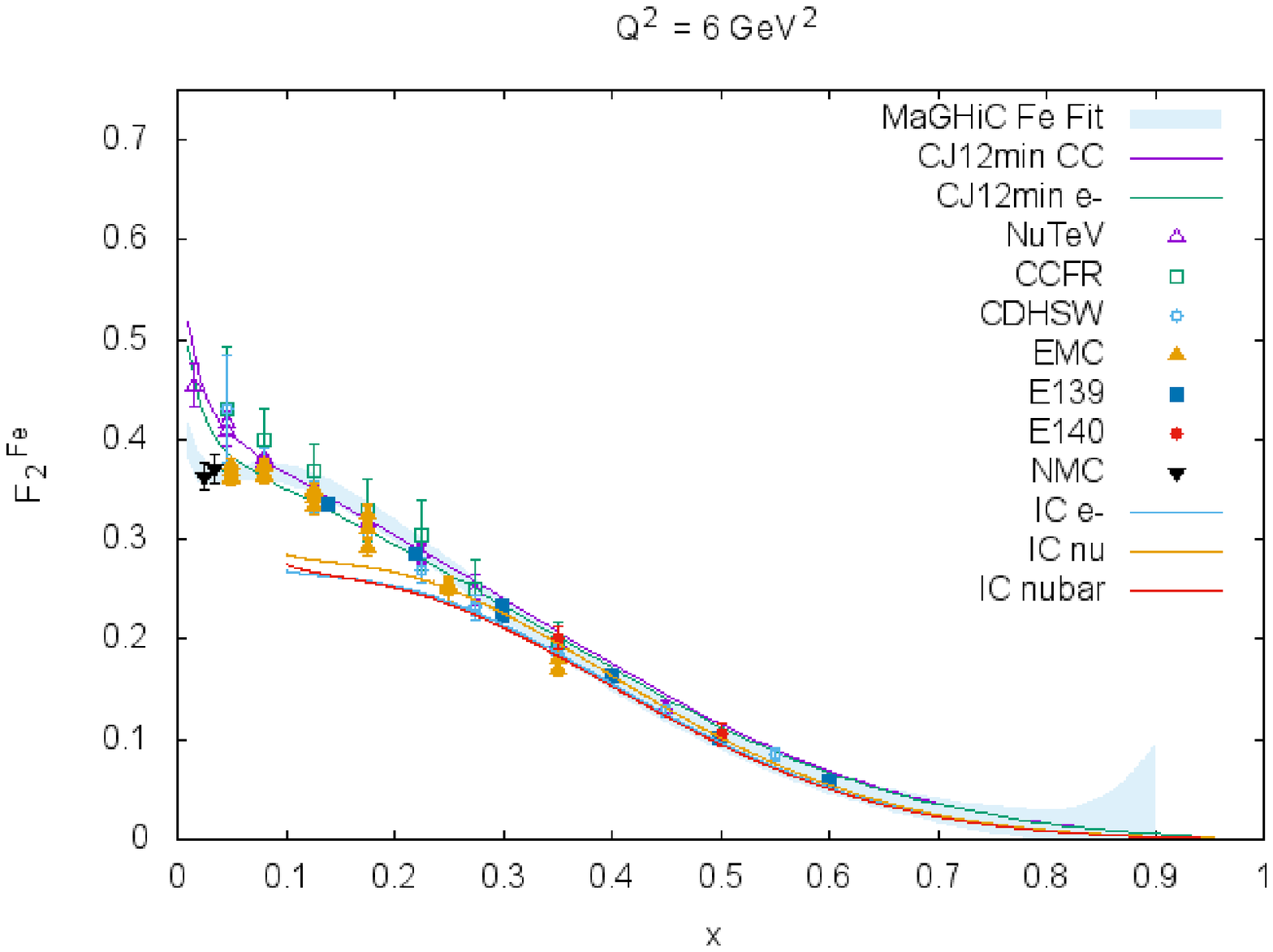}
\caption{\label{F2Fe_2-20_4-8} (Color online) $F_2^{Fe}$ data vs $x$. Data were obtained over $Q^2$ ranges of 2-20 GeV$^2$ (top) and 4-8 GeV$^2$ (bottom). The data and have been centered to a common $Q^2$ of 8 GeV$^2$ (top) and 6 GeV$^2$ (bottom) as described in the text. The curves also are as described in the text. }
\end{figure}

\noindent The data, centered and binned as described above, are shown versus x for different pre-centering bin sizes, as well as centered to different $Q^2$ values, in Figure \ref{F2Fe_2-20_4-8}. Different checks on $Q^2$ dependent and kinematic binning phenomena were performed. The $Q^2$ bin-centering correction was typically less than 5\%, with an evaluated model dependence of 4\% on this correction.

In all, there is remarkable agreement of the data sets within published uncertainties, which are typically only a few \%, above $x \approx 0.15$. This powerfully demonstrates the applicability of the 18/5 rule to nuclear data. While perhaps not surprising in the valence regime, it was however not a given. Higher twist effects or nuclear medium modifications, for instance, could have caused substantial deviations.


The data in Figure \ref{F2Fe_2-20_4-8} are compared to the CTEQ-JLab (CJ12) PDF ~\cite{CJ12}, to the MaGHiC nuclear ratio fit ~\cite{MaGHiC}, and to calculations made by Cloet et al ~\cite{IC1,IC2}. The CJ12 parameterization includes only deuterium nuclear corrections, and produces $F_2^N$ for the nucleons. From that $F_2^N$, $F_2^{Fe}$ was built here by adding the 26 neutrons and 30 protons. It was suggested ~\cite{AA} to use the CJmid option for the deuterium nuclear corrections used in the PDF extraction. The CJ12 fit was used for both the case of charged current neutrino (CC) scattering and electron scattering. For the CC case, neutrino and anti-neutrino were averaged for $F_2^N$ before constructing $F_2^{Fe}$. The CJ12 fit does not include nuclear effects beyond the deuteron, but it does take into account contributions from strange, and charm quarks for the neutrino's weak-coupling. This results in differing curves using the CJ fit for electrons and charged current neutrinos, providing a measure of the difference due to these effects.        

It may be surprising that the EMC effect, i.e. the nuclear dependence of the $F_2$ structure function in the region around $0.3 < x < 0.7$, is not visible in this larger $x$ regime when comparing the data to the CJ global fit. As a check, the $F_2^{Fe}$ data were divided by $F_2^D$ from the NMC parameterization, and were found to produce the expected EMC effect, which is simply too small to observe as plotted here rather than in the conventional ratio format. 

The MaGHiC curve shown on the Figures ~\cite{MaGHiC} is a parameterization of $F_2^A/F_2^D$ from charged lepton data, over a broad range of targets. To display $F_2^{Fe}$ here, the $F_2^A/F_2^D$ ratio from MaGHiC was multiplied by the NMC $F_2^D$ parameterization discussed above ~\cite{F2ALLM}. 

Also, included on the comparative plots are results from a calculation ~\cite{IC1,IC2} starting from a covariant quark Lagrangian, where no parameters are fit to structure function data. This model does not take anti-quarks or gluons into account. Hence, the observed undershooting at lower $x$ is expected, and the impressive agreement at higher $x$ is illustrative in showing what the contribution from the valence quarks may be.

While some reasonable concerns have periodically been raised in the literature about the analysis and consistency of the CDHSW and/or NuTeV data sets ~\cite{HPCS,BPZ}, we do not observe any significant discrepancy amongst the neutrino-Iron structure function data sets.

In contrast to the larger $x$ regime, the neutrino data are noticeably different from the charged lepton data in the lower $x$ region, $x < $ 0.15. The neutrino data seem rather consistent with the CJ12 curves, while the charged lepton data are in better agreement with the MaGHiC fit. The MaGHiC fit reflects the shadowing effects from a large set of charged lepton scattering data. CJ12, on the other hand, is garnered from only proton and deuteron data, with nuclear corrections applied to the latter to obtain free nucleon structure functions. The agreement of the neutrino data with the CJ nucleon therefore indicates a possible lack of nuclear medium modification to $F_2$ in the neutrino data, while not surprisingly the charged lepton data display the typical pattern of shadowing/anti-shadowing nuclear medium modifications. It is to be noted that MaGHiC encompasses also data from other nuclei where data are also available at lower $x$ than Fe, so that this charged lepton low $x$ behavior is well constrained. As noted previously, the difference of the two CJ12 curves demonstrates the magnitude of any difference caused by contributions from charm or strange quarks, which is a much smaller effect. The rather large observed difference between charged lepton and neutrino scattering data does not seem to have a significant $Q^2$ dependence, and persists also at higher $Q^2$ values which were studied. It becomes, however, increasingly difficult to visualize on the steeply rising low $x$ structure function curve.

\begin{figure}
\includegraphics[width=8cm,height=7cm]{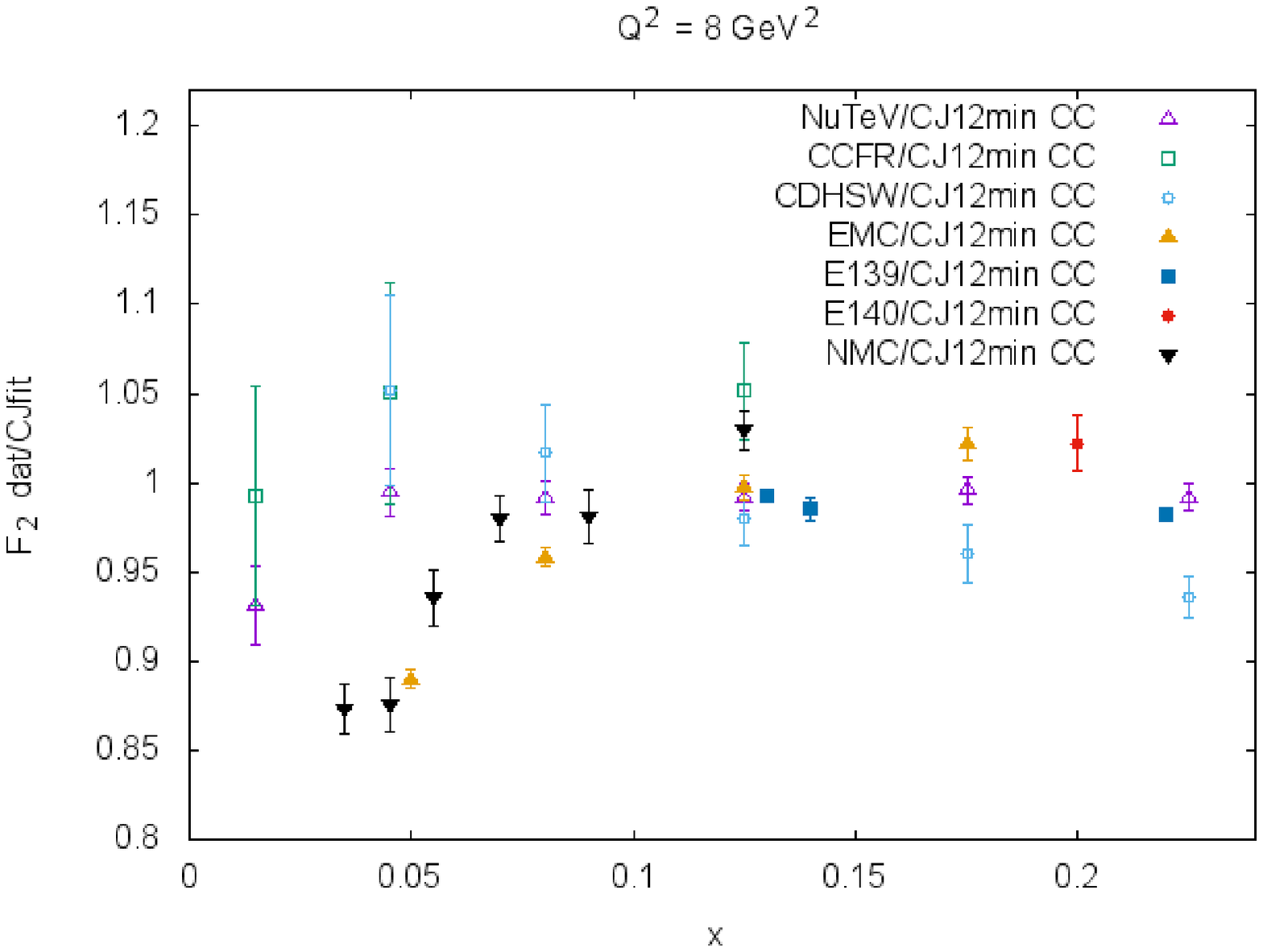}
\includegraphics[width=8cm,height=7cm]{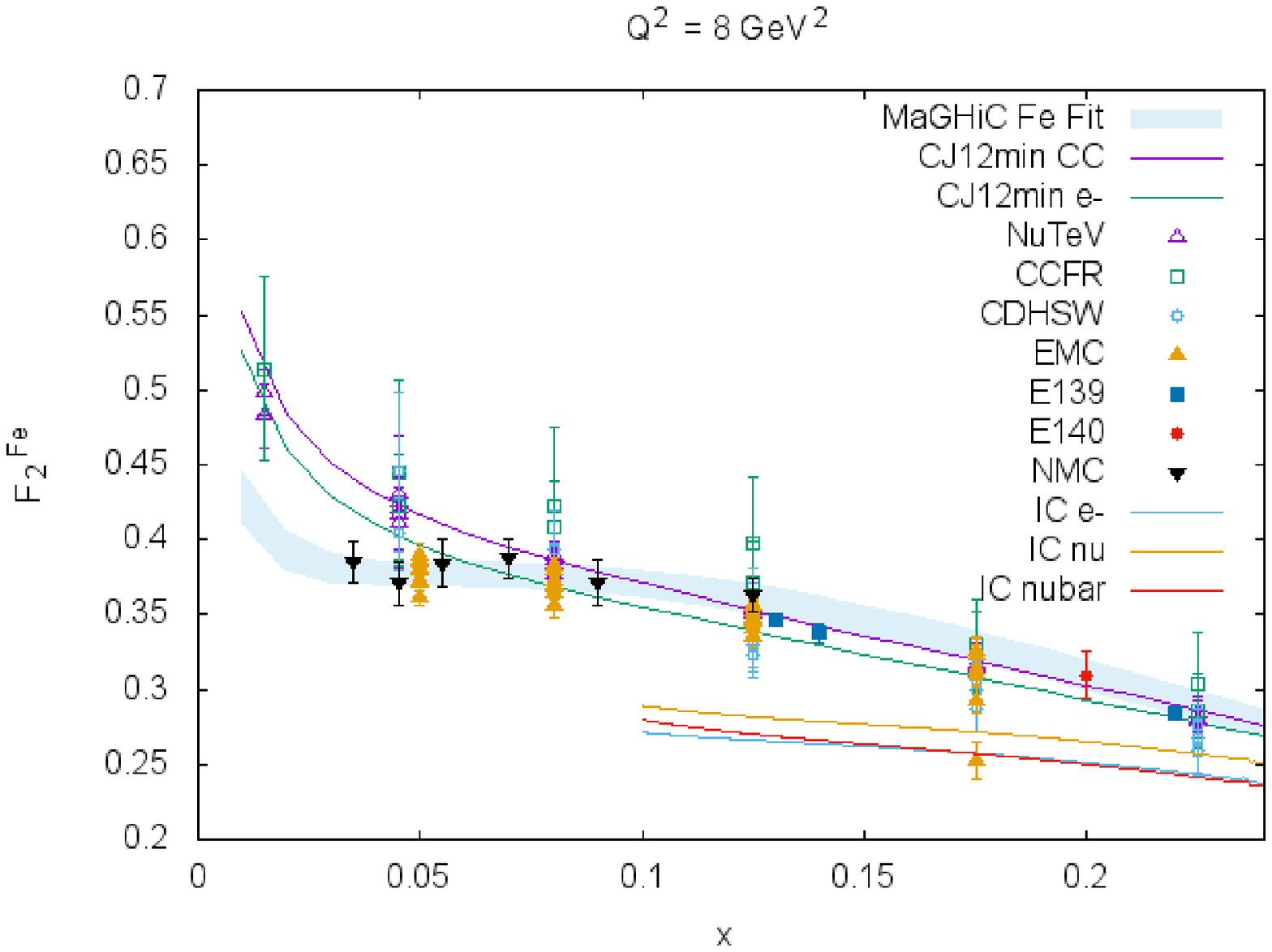}
\caption{\label{F2FeratCC_8} (Color online) (top) Ratio data/fit, of $F_2^{Fe}$ to the CJ12 fit for charged current neutrino vs the Bjorken scaling variable $x$, with (bottom) a low $x$ magnification of Figure~\ref{F2Fe_2-20_4-8} for $Q^2$ range of 2-20 GeV$^2$. In both cases, the data and fits are centered to a common $Q^2$ of 8 GeV$^2$. }
\end{figure}

To quantify the difference between the charged lepton and neutrino data at low $x$, we looked at the ratio, data/CJ, of the $F_2^{Fe}$ data to the CJ12 neutrino (anti-neutrino) $F_2$ fit. From the data/CJ ratio a difference of up to $\approx$ 15\% is observed between charged lepton and neutrino data. This can be seen in Figure ~\ref{F2FeratCC_8}. We also looked at the ratio of data/CJ electron fit where there is a small, 2-5\%, change - again providing some estimate of the strange quark contribution that is present in the neutrino case which is too small to account for the full observed effect. The neutrino and charged lepton scattering data consistently differ below $x < 0.15$, while agreeing well at larger $x$ values. The size of this observed difference is substantial in comparison for instance to the $\approx 5\%$ level EMC effect. 

Prevailing theories generally predict greater shadowing for the neutrino data. We observe in contrast the neutrino data to be consistenct with CJ, that is we observe reduced nuclear effects in the neutrino data as compared to the charged lepton data at low $x$. However, the data could alternatively be consistent with a general shift towards low $x$ of the medium modifications in neutrino data as sometimes predicted and also as observed by nCTEQ ~\cite{SK1,SK2,SK3}. In this case, shadowing may occur at somewhat lower $x$ for neutrino scattering as compared to charged lepton scattering and the CJ nucleon-only agreement would be rather accidental due to kinematic regime.

Recent results from the MINER$\nu$A ~\cite{MINERvA} neutrino scattering experiment appear to contradict the low $x$ data observation presented here. However, the MINER$\nu$A data are at low $Q^2$ and $W$ (also still somewhat preliminary at this time, and only available in nuclear ratios) and could be consistent with an $x$ shift of the data. Furthermore, it is not possible to directly compare our result presented here with the current MINERvA results, which are cross section ratios requiring inclusion as well of $xF_3$. The extended, higher energy MINERvA running for both neutrino and anti-neutrino will facilitate such a comparison.  

We note that the low $x$ nuclear charged lepton scattering data are dominated by a single experiment, NMC. Hence, the observations in this work are fully dependent on the accuracy of this data set. This will stay the case for some time as the currently available facilities can not achieve the energies to verify this data. The planned Electron-Ion Collider ~\cite{EIC}, however, can both verify and extend the range of the NMC experiment, while also providing both neutral and charged current lepton-nuclear scattering. It will be an ideal tool to further investigate the observations presented here. 

In summary, we have compiled and compared the world data for the Iron structure function $F_2^{Fe}$ within the DIS kinematic range $Q^2 > 2$ GeV$^2$ and $W > 4$ GeV$^2$, from both charged lepton and neutrino scattering data. There is remarkable agreement of all data using 18/5 scaling alone, also with available fits and calculations, in the valence region. We observe a substantial discrepancy, however, between the two types of data in the lower $x$ anti-shadowing and shadowing region. The discrepancy is on the order of 15\%, which is beyond what can be reasonably attributed to data or isoscalar correction uncertainties, or strange quark contributions. The observation is indicative that neutrino probes of nucleon structure might be sensitive to different nuclear effects than charged lepton probes at low $x$.

\section{Acknowledgments}
We thank A. Accardi for valuable discussions, and assistance with the CJ12 fitting.
We thank S. Malace for valuable discussions, and providing the MaGHiC fit.
We thank I. Cloet for providing his state-of-the-art calculations.  
We thank I. Schienbein and J. Morfin for valuable discussions.
This work was supported in part by 
research grant 1508272 from the National Science Foundation.
This material is based upon work supported by the U.S. Department of Energy, Office of Science, Office of Nuclear Physics under contract DE-AC05-06OR23177.


\bibliographystyle{apsrev4-1}
\bibliography{./F2Feccnc_nearfin-arxiv}



\end{document}